\documentclass[aps,prl,twocolumn]{revtex4}
\usepackage{amsmath}
\usepackage{graphicx}
\usepackage{multirow}
\usepackage{array}
\usepackage{inputenc}
\newcommand{\fecr}{Fe$_{1-x}$Cr$_x$}
\begin{document}
%
\author{I. Mirebeau}
\email[]{isabelle.mirebeau@cea.fr}
\author{G. Parette}
\affiliation{CEA, Centre de Saclay, DSM/IRAMIS/Laboratoire L\'eon Brillouin,
91191 Gif-sur-Yvette, France}
%

  \title{A new neutron study of  the short range order inversion in  Fe$_{1-x}$Cr$_x$}
  \date{\today}

  \begin{abstract}
We have performed new neutron diffuse scattering measurements in
Fe$_{1-x}$Cr$_x$ solid solutions, in a concentration range
0$<$x$<$0.15, where the atomic distribution shows an inversion of
the short range order. By optimizing the signal-background ratio,
we obtain an accurate determination of the concentration of
inversion x$_0$ =0.110(5). We determine the near neighbor atomic short
range order parameters and pair potentials, which
change sign at x$_0$. The experimental results are compared with previous first principle calculations and atomistic simulations.

\end{abstract}

\maketitle
\section{Introduction}
 FeCr alloys are a unique case in nature showing an inversion of the short range atomic order within a solid solution. Namely Cr atoms in a Fe matrix repel at low concentration, whether they attract at high concentration \cite{Mirebeau84}. This feature seems to be at the origin of the peculiar mechanical properties shown by FeCr alloys in the body centered cubic (bcc) solid solution, which have important potential applications. There is now a considerable effort in the search of materials highly resistant to radiation damages, which will occur both in future hybrid-types reactors and in controlled fusion reactors \cite{Cook06}. FeCr alloys with  body centered cubic (bcc) structures are the reference model to understand the behavior of ferritic FeCr steels, considered as leading candidates in most future nuclear energy options \cite{Klueh07}. Chromium substitution strongly changes the mechanical properties of iron, such as the swelling and formation of voids under irradiation, the radiation induced hardening, or the ductile- brittle transition. As a striking fact, the response of FeCr alloys under irradiation, as well as most of the mechanical properties,
 is highly non monotonic versus the Cr concentration, with a pronounced change of behavior around x=0.1.

 Understanding this behavior requires a detailed knowledge of the microscopic interactions between atoms, which govern the occupancy of the lattice by atoms of different kinds. In a solid solution of a binary alloy, the atomic distribution is never fully random, and short range ordered structures are stabilized in a statistical way, with preferential occupations controlled by the electronic configuration. In transition metal alloys, it has been known for decades that the electronic configurational energy  can be written in terms of effective pair potentials between atoms \cite{Ducastelle76,Ducastelle80,Bieber81}, although the total cohesive energy cannot \cite{Friedel69}.

The origin of the short range order (SRO) inversion in FeCr comes
from the electronic band structure. Introducing Cr atoms in the Fe
matrix leads to a lowering of the density of states at the Fermi
level, due to the formation of a virtual bounded state
\cite{Friedel69}. As a result, the effective pair potential
strongly varies with Cr concentration, and even changes sign at a
given Cr content. The band structure calculations outline the
dominant role played by Fe and Cr magnetism in the occurrence of
the inversion. In FeCr, the short range order inversion was first
of all predicted in the 80$^{ies}$ by $\it{ab}$  $\it{initio}$
calculations of M. Hennion \cite{Hennion83}, who extended to
ferromagnetic alloys the models developed in the Coherent
Potential Approximation (CPA) for paramagnetic transition metal
alloys \cite{Ducastelle76,Ducastelle80,Bieber81}. It was soon
after observed using resistivity and neutron probes
\cite{Mirebeau84}.
 More recently, motivated by the important applications in nuclear industry, several
 investigations of this effect were done both theoretically and experimentally. The SRO inversion
 was studied using M\"ossbauer spectroscopy \cite{Filipova00},
 and X-ray absorption fine structure (EXAFS) \cite{Froideval07}.
The mixing enthalpy was found to change sign around x=0.10
\cite{Wallenius04}. Many recent calculations of the pair
potentials in FeCr were also made, using either atomistic
simulations \cite{Caro05,Lavrentiev07}, first principle
thermodynamical calculations
\cite{Olsson05,Klaver06,Olsson07,Erhart08,Malerba08},
 or $\it{ab}$
$\it{initio}$ perturbation theories
\cite{Ruban08,Korzhavyi09,Turchi94}.


In this context, a precise determination of the pair potentials, as well as the concentration value where the short range order inversion occurs, is important. Neutron diffuse scattering is the best way to measure it, since the change of SRO directly affects the shape of the neutron cross section. This measurement is however difficult due to the small contrast between Fe and Cr neutron scattering lengths and to the low concentration range where the inversion occurs.
We have performed neutron measurements in \fecr\ alloys in the concentration range  0$<$x$<$0.15. With respect to our previous measurements \cite{Mirebeau84}, the experimental set-up was optimized, resulting in an increase of the signal over background ratio by more than an order of magnitude. We could therefore measure samples down to very low Cr contents (x= 0.01) and scan the inversion region carefully. These improved measurements yield a more precise determination of the short range order parameters, making possible to evaluate the pair potentials up to the fifth neighbors, and to localize the critical concentration for the short range order inversion accurately. The pair potentials are compared with theoretical determinations.

\section{Experimental conditions}
  Fifteen polycrystalline samples were prepared with concentrations in the range 0$<$x$<$0.15.
  They were synthesized by 
  CECM-Vitry (V), Cristaltech Grenoble (G), Gero-Neuhausen (L) and Cerac-Milwaukee (C), then
   shaped into cylinders of 30 mm length and 9 mm diameter. The samples homogeneity and the Cr
   concentrations were determined by chemical analysis. In the Fe-Cr system,
   the phase diagram shows a miscibility gap \cite{phasediagram} so that with increasing Cr concentration, the solid solution starts to decompose into two bcc phases, enriched in iron and chromium respectively.
   Although such decomposition may in principle occur in the concentration range studied here, its kinetics is much slower than the kinetics of SRO, which was precisely determined by measuring the residual resistivity \cite{Balanzat81,Pierron84,Mirebeau84}. The measurement of the self-diffusion coefficient, which follows an Arrhenius law with activation energy E=2.4 eV for a pre-exponential
  factor of 10$^{-14}$ s, determines a suitable heat treatment.

  After an homogenization at 800$^\circ$C, the samples were heated in a quartz tube at 520$^\circ$C, a
   temperature where an equilibrium state of SRO is immediately reached.
   Then the temperature T was gradually decreased down to 430$^\circ$C.
   The samples were kept at 430$^\circ$C for a few hours and quenched into water.
   This procedure permits to reach a stable SRO state at 430$^\circ$C and to
   preserve it during the quenching. 
   For the highest concentration (x=0.15),
   where the bcc solid solution may start to decompose at 430$^\circ$C,
   we studied the evolution of the SRO when the sample was annealed for much longer times.

   Neutron measurements were performed on the G6.1 diffractometer of the ORPHEE reactor in Saclay,
  with a incident neutron beam of 
   4.73 \AA\ wavelength provided by a focusing graphite monochromator.
   The range of the scattering vector (0.1$<$K$<$ 2.5 \AA$^{-1}$), allowed measuring the diffuse
   scattering with a good accuracy without any contribution from the Bragg scattering.
  To decrease the contribution from inelastic scattering, the measurements were
   performed at low temperature (8~K) using a cryogenerator. A vertical magnetic field of 15 kOe
   provided by an electromagnet was applied to saturate the sample in the direction
   perpendicular to the scattering plane ($\bf{K}$$\perp$$\bf{H}$). Combined with zero field measurements,
   this procedure is used to separate the nuclear and magnetic cross sections \cite{Cable}.
   Both electromagnet and cryogenerator were placed inside a vacuum chamber,
   which decreased the environmental background
   by a factor 5.
   Combining this with the higher neutron flux due to high
   intensity neutron beam and focusing monochromator,
   the signal/background ratio was improved by a factor 12 with respect to that of ref. \cite{Mirebeau84}.
  The neutron intensities were corrected for background, absorption and multiple scattering, and calibrated in absolute scale by measuring a vanadium standard.
\section{Results and analysis}

In the nuclear diffuse cross section of a FeCr dilute alloy, the contribution of static lattice distortions can be neglected, since the atomic radii of Fe and Cr are very similar. The nuclear cross section then reduces to the incoherent cross section and contribution from atomic short range order. It is shown in Fig. 1 as a function of the scattering vector K, for most samples in the region of the SRO inversion, and in Fig. 2 for samples with the highest Cr content x=0.15. For x =0, the nuclear cross section is K-independent and equal to the incoherent cross section of iron. For 0$<$x$<$0.10, it clearly decreases at small K values, showing a tendency to short range order. For x=0.107, the cross section is K-independent again, which shows that the atomic distribution is random. Finally,
for x=0.117 and above, the strong increase of the cross section at small K values shows the presence of short range clustering.

Following the model of Cowley-Warren \cite{Cowley50}, the
nuclear cross section for a binary alloy Fe$_{1-x}$Cr$_{x}$ is expressed as:
 \begin {equation}
\frac{d\sigma}{d\Omega}=\frac{\sigma_{inc}}{4 \pi} + x(1-x)(b_{Fe}-b_{Cr})^2 S(K),
 \label{neutroncross}
\end{equation}

 where $\sigma$$_{inc}$ is the incoherent nuclear cross section of the alloy: $\sigma$$_{inc}$  = (1-x) $\sigma$$_{inc}^{Fe}$ +x  $\sigma$$_{inc}^{Cr}$, with  $\sigma$$_{inc}^{Fe}$ = 0.427 and  $\sigma$$_{inc}^{Cr}$ = 2.538 barn. b$_{Fe}$ and b$_{Cr}$ are the  nuclear coherent scattering lengths: b$_{Fe}$ = 0.951 and b$_{Cr}$ =0.352 10$^{-12}$cm. The short range order
   function S(K), averaged for a polycrystalline alloy over all $\bf{K}$ orientations, writes:

 \begin{equation}
 S(K) = 1+ \sum_{n} z_n \alpha_n \frac{sin K R_n}{K R_n},
\label{spectral function}
\end{equation}
 where  z$_n$, R$_n$ and  $\alpha$$_n$ are respectively the coordination number, radius and SRO parameter of the n$^{th}$ shell surrounding an atom placed at the origin. A positive (resp. negative) $\alpha$$_n$  value corresponds to an attractive (resp. repulsive) type of order
 between the central atom and the atoms of the same species in the shell considered.

\begin{figure}
\begin{center}
  \includegraphics[width=8cm]{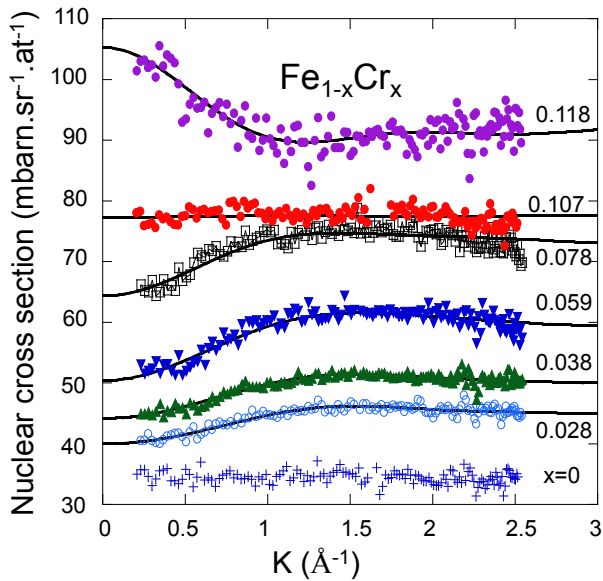}
  \caption{Nuclear neutron diffuse cross section of Fe$_{1-x}$Cr$_x$ alloys. Solid lines are fits with the Cowley-Warren model, involving SRO parameters up to the 5$^{th}$ shell.
    } \label{fig1}
\end{center}
\end{figure}

\begin{figure}
\begin{center}
  \includegraphics[width=8cm]{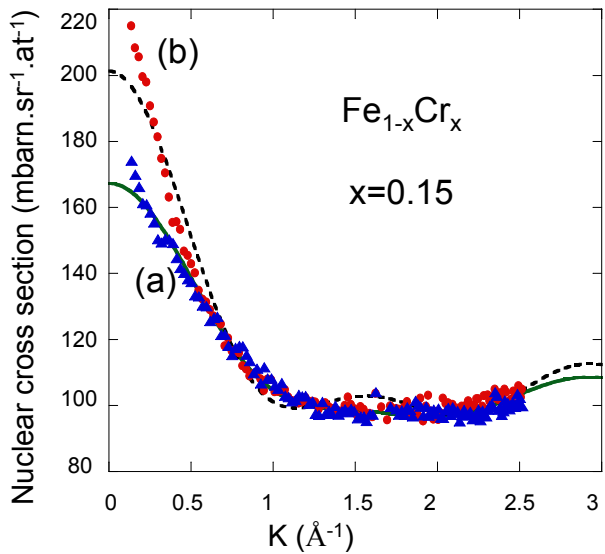}
  \caption{ Color on line. Nuclear neutron diffuse cross section of Fe$_{1-x}$Cr$_x$ alloys annealed at 430$^\circ$C for two annealing times: (a): x=0.152(1), annealing time 18 hours; triangles: experiment; solid line: fit. (b) x=0.151(1): annealing time 3.5 days; dots : experiment; dashed line: fit. Fits are performed with SRO parameters up to the 5$^{th}$ shell.
    } \label{fig2}
\end{center}
\end{figure}


Due to the limited K range and to the averaged information given by the polycrystalline samples,
 the number of parameters must be restricted to the first neighboring shells. Moreover,
 it was not possible to determine the SRO parameters on close concentric shells separately.
 For a bcc alloy, the shells 1 and 2 (R/a =0.866 and 1, where a is the lattice constant),
  4 and 5 (R/a =1.66 and 1.73),
 are close to each other,
 and we have grouped them in the analysis. The averaged SRO parameter corresponding to the grouped (i,j) shell of coordination z$_i$+z$_j$ is defined as
  $\alpha$$_{ij}$=(z$_i$$\alpha$$_i$+z$_j$$\alpha$$_j$)/(z$_i$+z$_j$). The experimental cross sections
  were fitted according to the above equations with the SRO parameters as fitting parameters.
  We obtain a good fit of the experimental data for all samples
  as shown by solid curves in Figs. 1 and 2.
  This  procedure allowed an accurate determination of the SRO parameters for
  all samples. For x=0, we found an absolute value of the nuclear cross section in perfect agreement with the calculated incoherent cross section of iron (32(2) mbarn.sr$^{-1}$.at$^{-1}$), which ensures that all calibrations and corrections have been done properly. The SRO parameters are shown in Fig. 3 versus concentration.
  As expected the SRO parameters decrease when increasing  the size of the coordination shell.
  The inversion of short range order is clearly seen on the $\alpha$$_{12}$  parameter,
  which has the strongest value. At low concentrations  $\alpha$$_{12}$ is negative and
  close to the curve -x/(1-x) which corresponds to the maximum repulsion between Cr atoms.
  With increasing concentration, it
   changes sign 
  and becomes strongly positive, as expected for a tendency to segregation. The SRO parameter for the third atomic shell $\alpha$$_3$ is much smaller, but shows the same tendencies. In the 4-5 shells the SRO parameter is concentration independent and close to zero, which corresponds to a random atomic distribution. The plot of the extrapolated value of the correlation function S(0)=  $\sum_{i} z_i \alpha_i$  versus x (Fig. 3), determines the inversion concentration as x$_0$=0.110(5) for which S(0)=1. As noticed earlier \cite{Clapp66}, S(0) must be zero in a canonical ensemble, but this infinitely narrow singularity in S(k) should never be observed experimentally.
 For the sample with the highest Cr content which belongs to the region of the miscibility gap at 430$^\circ$C,
 we compare in Fig. 2
 the nuclear cross section of samples annealed at long annealing times (18h and 3.5 days). The decomposition of the alloys in Cr- rich clusters starts to occur, as shown by the pronounced enhancement of the nuclear cross section at low K values.
 This enhancement cannot be correctly accounted for by the Cowley-Warren model, when only a few SRO parameters are involved.
 It becomes more pronounced as the annealing time increases.
\begin{figure}
\begin{center}
        \includegraphics[width=8cm]{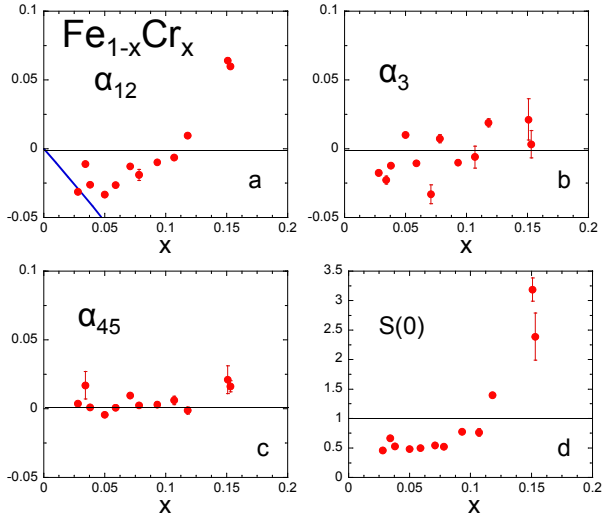}
  \caption{ Color on line. Variation of the SRO parameters $\alpha$$_i$ with concentration x in Fe$_{1-x}$Cr$_x$ alloys. a) average parameter for the first two neighbor shells; the solid line is the curve for maximal repulsion between Cr nearest and next nearest neighbors; b) third shell; c) 4-5 shells; d) calculated value of the short range ordered function at k=0.
    } \label{fig3}
\end{center}
\end{figure}

 Knowing the SRO parameters for a well-defined temperature state, one can deduce the pair interaction potentials
 in the Krivoglaz-Clapp-Moss mean field approach \cite{Clapp66,Krivoglaz}.
In the high temperature limit,  the correlation function S(K) is related to the Fourier transform V(K) of the pair potential through the expression:
\begin{equation}
 S(K) =  \frac{1}{1+2 x(1-x) \frac{V(K)}{k_\beta T_0}},
\label{potential}
\end{equation}

where T$_0$=430$^\circ$C=703~K is the temperature for the quench-in state of the alloys.
  Eq. \ref{potential} can be linearized to obtain the
 pair potentials V$_i$ for the first neighbor shells.
  These pair potentials are shown in Fig. 4 versus concentration.
  As expected from the general behavior in transition metal alloys, for a given concentration,
  the pair potentials decrease when increasing the size of the coordination shell.
  Here,
  one notices that the three potentials seem to change sign at about the same concentration.
\begin{figure}
\begin{center}
        \includegraphics[width=8cm]{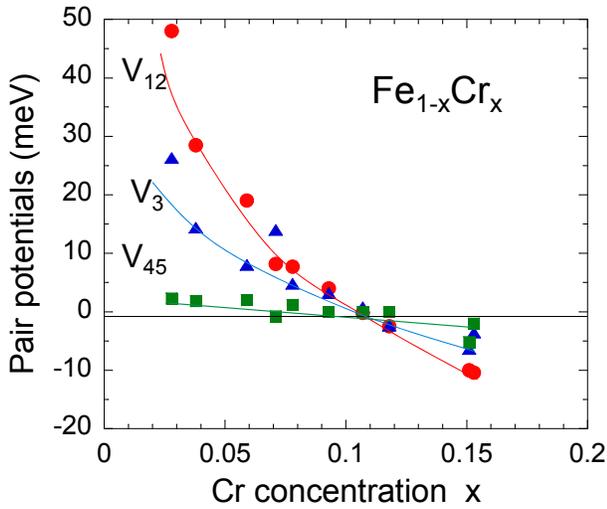}
  \caption{Colour on line. Variation of the pair potentials V$_i$ with concentration x in Fe$_{1-x}$Cr$_x$ alloys. Solid lines are guides to the eye.
    } \label{fig4}
\end{center}
\end{figure}

\section{Discussion}
 The knowledge of the first neighbor pair potential is important, since it is the main ingredient to predict the FeCr phase diagram \cite{Caro06,phasediagram,Mathon03}, in a range of concentration and temperature where it cannot be measured at equilibrium, but which is  important for nuclear applications. The concentration range where the atomic distribution is almost random allows one to limit the segregation process induced by irradiation, keeping the mechanical properties induced by Cr substitution.
\begin{figure}
\begin{center}
        \includegraphics[width=8cm]{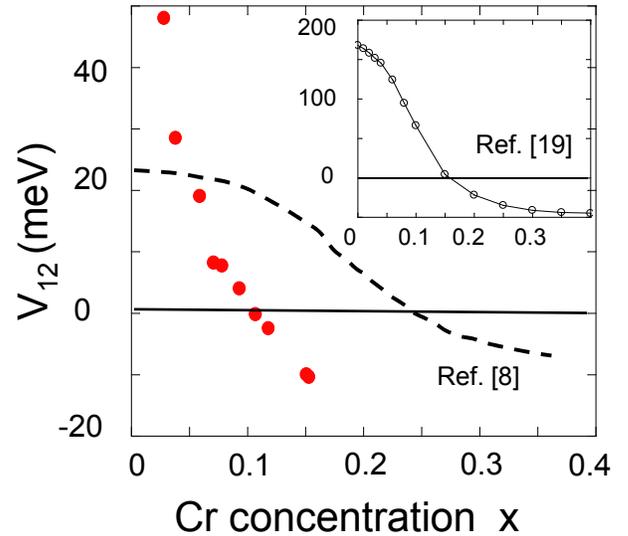}
  \caption{Colour on line. 
    Variation of the Pair potential V$_{12}$  for the first two shells with Cr concentration x; (red dots): this experiment; dashed line: $\it{ab}$ $\it{initio}$ calculation of ref.\cite{Hennion83}; Inset: calculated potential deduced from ref.\cite{Ruban08}. The solid line in the inset is a guide to the eye.
    } \label{fig4}
\end{center}
\end{figure}

In Fig. 5, we compare the variation of
V$_{12}$ 
versus concentration with
theoretical evaluations.
 In the $\it{ab}$  $\it{initio}$ calculation of \cite{Hennion83},
the inversion for short range ordering is predicted for x$_0$=0.25
instead of 0.11. In spite of this discrepancy, the quantitative
agreement about the potential V$_{12}$ is remarkable, considering
that the energy of configuration represents a very small
 part of the total cohesive energy.

 The Cr impurity yields a strong deformation of the up band,
 with a localized state above the Fermi level and a minimum of the density of states
 below. The occurrence of a negative moment on the Cr site (namely antiparallel to the Fe
 moment) and the change of sign of the pair potential $\it{versus}$ Cr content are closely
 related, both being the consequences of this deformation. Recent first principle calculations,
 using the screened generalized perturbation method \cite{Ruban08,Korzhavyi09}
 confirm the dramatic changes of the Fermi surface topology
 in the majority spin channel of  FeCr alloys, found to occur
 in a narrow concentration range between 0.05 and 0.10. As a result, the first four pair potentials
 decrease abruptly with increasing Cr content.  The potential V$_{1}$ is found to change sign at x$\sim$ 0.15, and its concentration dependence mimics that of the local Cr
moment. The potential V$_{12}$, evaluated as the weighted average of the first two shells
(V$_{12}$=(8V$_{1}$+6V$_{2}$)/14),  is however larger than
the experimental one (inset fig. 5).
 As argued in ref \cite{Ruban08}, to compare with experiment, one should take into account the influence
  of the magnetic state of the alloy during the annealing. In
  contrast,
 calculations made in the non magnetic or paramagnetic state \cite{Korzhavyi09, Turchi94}
 predict a much smoother variation of the first pair potential.

 In real
 space, the repulsion of diluted Cr atoms in the Fe matrix
 may be understood as a magnetic frustration effect, since Cr
 moments are antiferromagnetic (AF) in pure Cr and in the bcc Cr-rich phase.
 In nearest neighbor Cr-Cr pairs, the Cr moments should be parallel to be AF coupled
 with the surrounding Fe moments, the Cr-Cr AF coupling would then be frustrated.
 Our
 measurements of the magnetic cross section of FeCr alloys
(to be detailed later)
 confirm the existence of a negative moment of
 -0.8(1) $\mu$$_B$ on the Cr site, in rather good agreement with previous experimental
results \cite{Collins64,Campbell66,Aldred76} and theoretical determinations \cite{Hennion83,Klaver06}. At low Cr content (x$\sim$0.01-0.03), the magnetic cross section, which reflects the perturbation induced by Cr moments on
 neighboring Fe sites (determined from the magnetic SRO parameters) is quite close to that calculated for an
 isolated Cr moment in the Fe matrix \cite{Drittler89}.

 In Fig. 6, we compare our experimental results to the variation of the
 SRO parameter $\alpha$$_{12}$ calculated by evaluating the mixing enthalpy.
 In ref.\cite{Caro05}, a method was proposed to generalize classical many body potentials,
 which could predict the inversion of the pair potential and change of sign in the mixing enthalpy.
  Further Monte-Carlo simulations were performed within this approach, and investigated the influence of the heat treatment on the SRO inversion, taking into account the miscibility gap and the formation of stable precipitates from the Cr-rich phase \cite{Erhart082}. At the equilibrium temperature of 700~K, almost the equilibrium temperature of our experiment (703~K), the concentration of inversion is evaluated as x$\sim$ 0.12, rather close to the experimental value.
  Another Monte-Carlo study of the thermodynamic properties of FeCr alloys
  is based on a cluster expansion of the configurational contributions to the mixing enthalpy \cite{Lavrentiev07}.
  It yields an inversion concentration x=0.105 at T=750~K 
  in very good agreement with the experimental value. We note however that in both cases the calculated $\alpha$$_{12}$ parameters are larger than the experimental ones, and have a different concentration dependency.

\begin{figure}
\begin{center}
        \includegraphics[width=8cm]{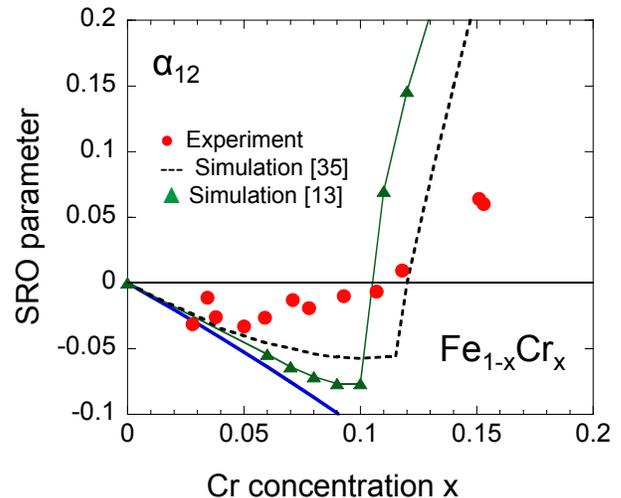}
  \caption{ Color on line SRO parameter $\alpha$$_{12}$ versus Cr concentration x; (dots): this experiment;
  dashed line: simulation from ref.\cite{Erhart082};
  (triangles): simulation from ref.\cite{Lavrentiev07}; the thick blue line is the maximum repulsion between Cr nearest and next nearest neighbors.
   The thin line is a guide to the eye.
    } \label{fig5}
\end{center}
\end{figure}

 To summarize, we studied the short range order inversion in Fe-Cr alloys by neutron scattering, with a much better accuracy than before. The concentration of inversion is found to be x$_0$=0.110(5). The pair potential and short range parameters are compared with recent calculations. This comparison provides a stringent test of the calculated pair potentials, which may be useful for future models and predictions of the FeCr phase diagram.

 We thank S. Michaud for her participation to the experiment and data treatment, F. Ducastelle for stimulating discussions and V. Pierron-Bohnes for a critical reading of the manuscript.

\end{document}